\documentclass[10pt,a4paper]{article}
\usepackage{amsmath,amssymb,latexsym}
\usepackage{mathrsfs,graphicx,subfigure}

\def\AFOUR{%
\setlength{\textheight}{8.5in}%
\setlength{\textwidth}{5.75in}%
\setlength{\topmargin}{-0.375in}%
\hoffset=-.5in%
\renewcommand{\baselinestretch}{1.17}%
\setlength{\parskip}{6pt plus 2pt}%
}

\AFOUR                                           

\expandafter\ifx\csname amssym.def\endcsname\relax \else\endinput\fi
\expandafter\edef\csname amssym.def\endcsname{%
       \catcode`\noexpand\@=\the\catcode`\@\space}
\catcode`\@=11
\def\undefine#1{\let#1\undefined}
\def\newsymbol#1#2#3#4#5{\let\next@\relax
 \ifnum#2=\@ne\let\next@\msafam@\else
 \ifnum#2=\tw@\let\next@\msbfam@\fi\fi
 \mathchardef#1="#3\next@#4#5}
\def\mathhexbox@#1#2#3{\relax
 \ifmmode\mathpalette{}{\m@th\mathchar"#1#2#3}%
 \else\leavevmode\hbox{$\m@th\mathchar"#1#2#3$}\fi}
\def\hexnumber@#1{\ifcase#1 0\or 1\or 2\or 3\or 4\or 5\or 6\or 7\or 8\or
 9\or A\or B\or C\or D\or E\or F\fi}

\font\tenmsa=msam10
\font\sevenmsa=msam7
\font\fivemsa=msam5
\newfam\msafam
\textfont\msafam=\tenmsa
\scriptfont\msafam=\sevenmsa
\scriptscriptfont\msafam=\fivemsa
\edef\msafam@{\hexnumber@\msafam}
\mathchardef\dabar@"0\msafam@39
\def\dashrightarrow{\mathrel{\dabar@\dabar@\mathchar"0\msafam@4B}}
\def\dashleftarrow{\mathrel{\mathchar"0\msafam@4C\dabar@\dabar@}}

\def\ulcorner{\delimiter"4\msafam@70\msafam@70 }
\def\urcorner{\delimiter"5\msafam@71\msafam@71 }
\def\llcorner{\delimiter"4\msafam@78\msafam@78 }
\def\lrcorner{\delimiter"5\msafam@79\msafam@79 }
\def\yen{{\mathhexbox@\msafam@55}}
\def\checkmark{{\mathhexbox@\msafam@58}}
\def\circledR{{\mathhexbox@\msafam@72}}
\def\maltese{{\mathhexbox@\msafam@7A}}
\def\circledS{{\mathhexbox@\msafam@73}}

\font\tenmsb=msbm10
\font\sevenmsb=msbm7
\font\fivemsb=msbm5
\newfam\msbfam
\textfont\msbfam=\tenmsb
\scriptfont\msbfam=\sevenmsb
\scriptscriptfont\msbfam=\fivemsb
\edef\msbfam@{\hexnumber@\msbfam}
\def\Bbb#1{{\fam\msbfam\relax#1}}
\def\widehat#1{\setbox\z@\hbox{$\m@th#1$}%
 \ifdim\wd\z@>\tw@ em\mathaccent"0\msbfam@5B{#1}%
 \else\mathaccent"0362{#1}\fi}
\def\widetilde#1{\setbox\z@\hbox{$\m@th#1$}%
 \ifdim\wd\z@>\tw@ em\mathaccent"0\msbfam@5D{#1}%
 \else\mathaccent"0365{#1}\fi}
\font\teneufm=eufm10
\font\seveneufm=eufm7
\font\fiveeufm=eufm5
\newfam\eufmfam
\textfont\eufmfam=\teneufm
\scriptfont\eufmfam=\seveneufm
\scriptscriptfont\eufmfam=\fiveeufm
\def\frak#1{{\fam\eufmfam\relax#1}}

\csname amssym.def\endcsname

\makeatletter
\def\section{\@startsection {section}{1}{\z@}{-3.5ex plus -1ex minus
 -.2ex}{2.3ex plus .2ex}{\large\sc}}
\def\subsection{\@startsection{subsection}{2}{\z@}{-3.25ex plus -1ex minus
 -.2ex}{1.5ex plus .2ex}{\normalsize\sc}}
\makeatother

\makeatletter
\@addtoreset{equation}{section}

\makeatother

\newcommand{\nc}{\newcommand}
\newcommand{\rnc}{\renewcommand}

\nc{\chap}[1]{{\clearpage}%
\begin{center}%
{\noindent\underline{\large\sc #1}}{\addcontentsline{toc}{section}{#1}}%
\end{center}%
{\vspace*{0.3cm}}}

\nc{\subs}[1]{{\vspace*{0.2cm}}%
{\noindent\underline{\small\sc
#1}}%
{\vspace*{0.2cm}}}

\nc{\be}{\begin{equation}}
\nc{\ee}{\end{equation}}
\nc{\bea}{\begin{eqnarray}}
\nc{\eea}{\end{eqnarray}}

\nc{\trac}[2]{{\textstyle\frac{#1}{#2}}}

\nc{\ex}[1]{\mbox{e}^{\,\textstyle#1}}

\nc{\CC}{\Bbb{C}}
\nc{\HH}{\Bbb{H}}
\nc{\PP}{\Bbb{P}}
\nc{\RR}{\Bbb{R}}
\nc{\ZZ}{\Bbb{Z}}
\nc{\II}{\Bbb{I}}
\nc{\EE}{\Bbb{E}}
\nc{\TT}{\Bbb{T}}
\nc{\DD}{\mathrm{I}\!\mathrm{D}}

\rnc{\d}{\delta}
\nc{\eps}{\epsilon}
\nc{\om}{\omega}

\nc{\symx}{\circledS}
\newsymbol\smallsmile 1360
\newsymbol\smallfrown 1361
\nc{\ad}{\mathop{\mbox{ad}}\nolimits}
\nc{\tr}{\mathop{\mbox{tr}}\nolimits}
\nc{\Tr}{\mathop{\mbox{Tr}}\nolimits}
\nc{\Det}{\mathop{\mbox{Det}}\nolimits}
\rnc{\det}{\mathop{\mbox{det}}\nolimits}
\nc{\rk}{\mathop{\mbox{rk}}\nolimits}
\nc{\del}{\partial}
\nc{\diag}{\mathop{\mbox{diag}}\nolimits}
\nc{\ra}{\rightarrow}
\nc{\Ra}{\Rightarrow}
\nc{\LRa}{\Leftrightarrow}
\nc{\lra}{\leftrightarrow}
\nc{\ot}{\otimes}
\rnc{\ss}{\subset}
\nc{\nul}{\noindent\underline}
\nc{\non}{\nonumber\\}
\nc{\mat}[4]{\left(\begin{array}{cc}#1&#2\\#3&#4\end{array}\right)}
\rnc{\lg}{\frak{g}}
\nc{\G}[3]{\Gamma^{#1}_{\;{#2}{#3}}}
\nc{\nam}{\nabla_{\mu}}
\nc{\nan}{\nabla_{\nu}}
\nc{\dx}{\dot{x}}
\nc{\tx}{\tilde{x}}
\nc{\dtx}{\dot{\tilde{x}}}
\nc{\te}{\tilde{e}}
\nc{\dte}{\dot{\tilde{e}}}
\nc{\dxl}{\dot{x}^{\la}}
\nc{\dxm}{\dot{x}^{\mu}}
\nc{\dxn}{\dot{x}^{\nu}}
\nc{\ddx}{\ddot{x}}
\nc{\ddxm}{\ddot{x}^{\mu}}
\nc{\ddxn}{\ddot{x}^{\nu}}
\nc{\dxi}{\dot{\xi}}
\nc{\ddxi}{\ddot{\xi}}
\nc{\lsf}{\ell_s^{\mathrm{eff}}}
\nc{\lpf}{\ell_p^{\mathrm{eff}}}
\nc{\sqg}{\sqrt{g^{11}}}

\nc{\bpm}{\begin{pmatrix}}
\nc{\epm}{\end{pmatrix}}

\nc{\red}[1]{{\color{red}#1}}
\nc{\dd}{\mathrm{d}}

\begin{document}
\rightline{Version of \today}

\begin{center}
{\Large\sc A linear mass Vaidya metric at \\
the end of black hole evaporation}
\end{center}

\vspace{.5cm}

\begin{center}
{\large\sc Martin O'Loughlin}\\[.4cm]
{\it University of Nova Gorica, Vipavska 13, 5000 Nova Gorica,
Slovenia}\\[.3cm]
\end{center}

\begin{abstract}
We discuss the near singularity region of the linear mass Vaidya metric for 
massless particles with non-zero angular momentum.  In particular we look
at massless geodesics with non-zero angular momentum near the vanishing 
point of a special subclass of linear mass Vaidya metrics. We also investigate
this same structure
in the numerical solutions for the scattering of massless scalars 
from the singularity. Finally we make some comments
on the possibility of using this metric as a semi-classical model
for the end-point of black hole evaporation. 
\end{abstract}

\section{Introduction}

\begin{figure}[h]
\centering
\includegraphics[width=.65\columnwidth,keepaspectratio=]{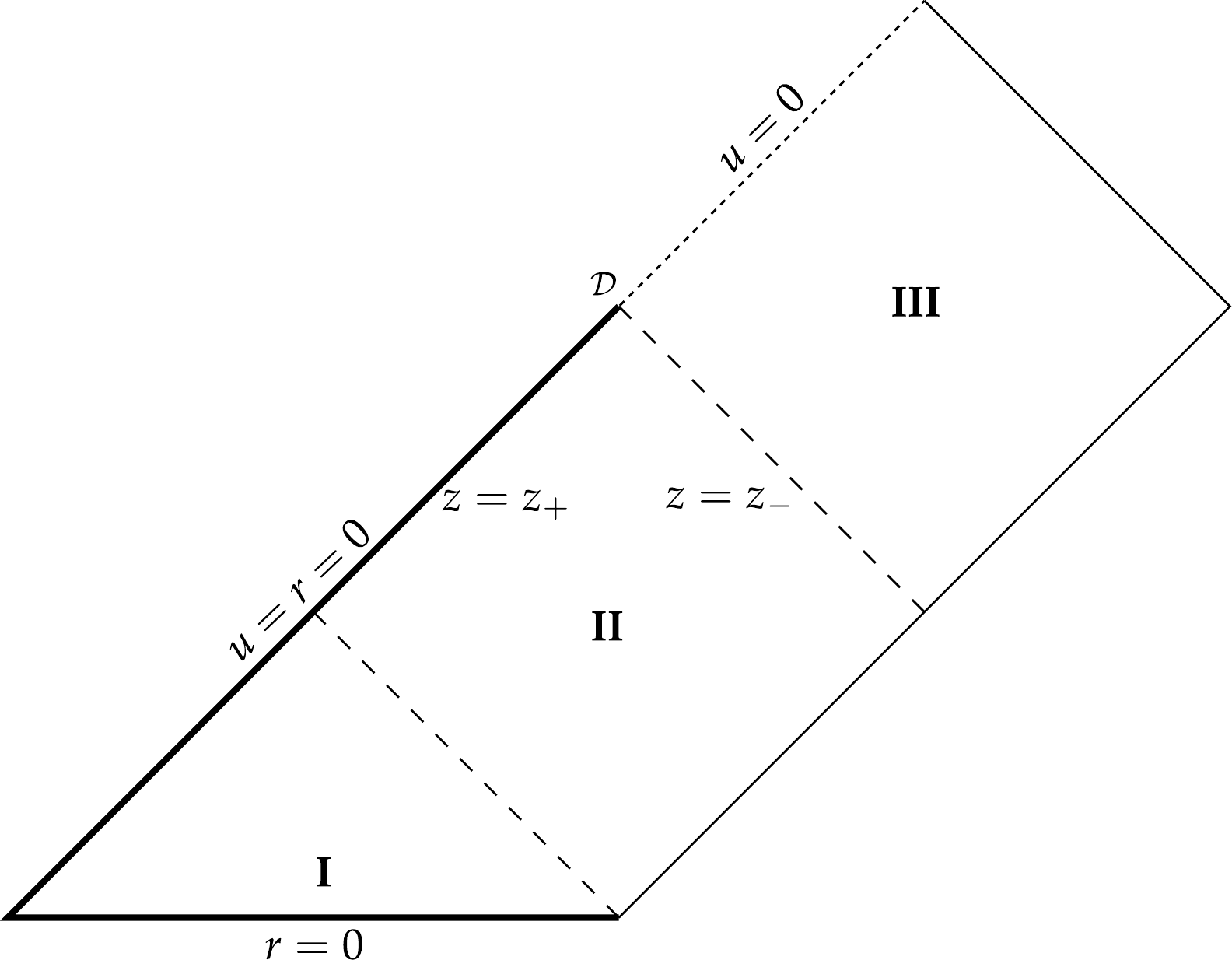}
\caption{Penrose diagram for the linear mass Vaidya metric with
$0<\mu<1/16$, $\mathcal{D}$ is the vanishing point.}
\label{fig:linearMvaidya}
\end{figure}

The linear mass Vaidya metric is a special class of Vaidya metrics
\cite{vaidya43,exactsolutions,vaidya51}  over 
which one has a certain degree of analytic control, 
in particular as a consequence of the additional homothety symmetry that 
these metrics possess. In general the Vaidya metric has the form
\begin{equation}
\dd s^2 =
  -\left(
    1 - \frac{2m(u)}{r}
  \right)\dd u^2 -
  2\,\dd u\,\dd r +
  r^2\dd\Omega^2
\end{equation}
and when $m(u) = -\mu u$ this metric has a homothety symmetry under
rescaling of the coordinates $u$ and $r$ together with an 
overall rescaling of the metric. In addition, when $0<\mu<1/16$ these metrics
have the special property that they contain a null singularity that vanishes
at an interior point of the spacetime. These metrics have been studied
in detail in \cite{waugh861} and for the above range 
of $\mu$ the detailed causal structure of the space-time is 
illustrated in the Penrose diagram of figure (\ref{fig:linearMvaidya}) where
$\mathcal{D}$ indicates the vanishing point. These metrics have been used 
in \cite{waugh862,bicak97,hiscock82,kuroda,balbinot,abdalla,bicak03,ghosh,
harko,girotto,kawai,fayos10,farley} 
to study various aspects of black hole evaporation and cosmic censorship.
Nevertheless there is still no consistent picture of the complete evaporation
of a black hole. The current paper does not claim to resolve this question
but presents various arguments to support the proposal that the final stages 
of evaporation can be modelled at the semi-classical level by a linear
mass Vaidya metric.

The outgoing Vaidya metric has also recently been used in \cite{lowe} to 
study evaporating black-holes and unitarity. Questions of information loss
and unitarity are of great interest but the current paper does not address
them as we
are specifically looking at the evolution from the Page time \cite{page} 
(the time after which, assuming that evaporation is 
unitary, a black hole has reemitted almost all of the information 
that it contained), up to its complete evaporation.

\begin{figure}
\centering
\includegraphics[width=.55\columnwidth,keepaspectratio=]{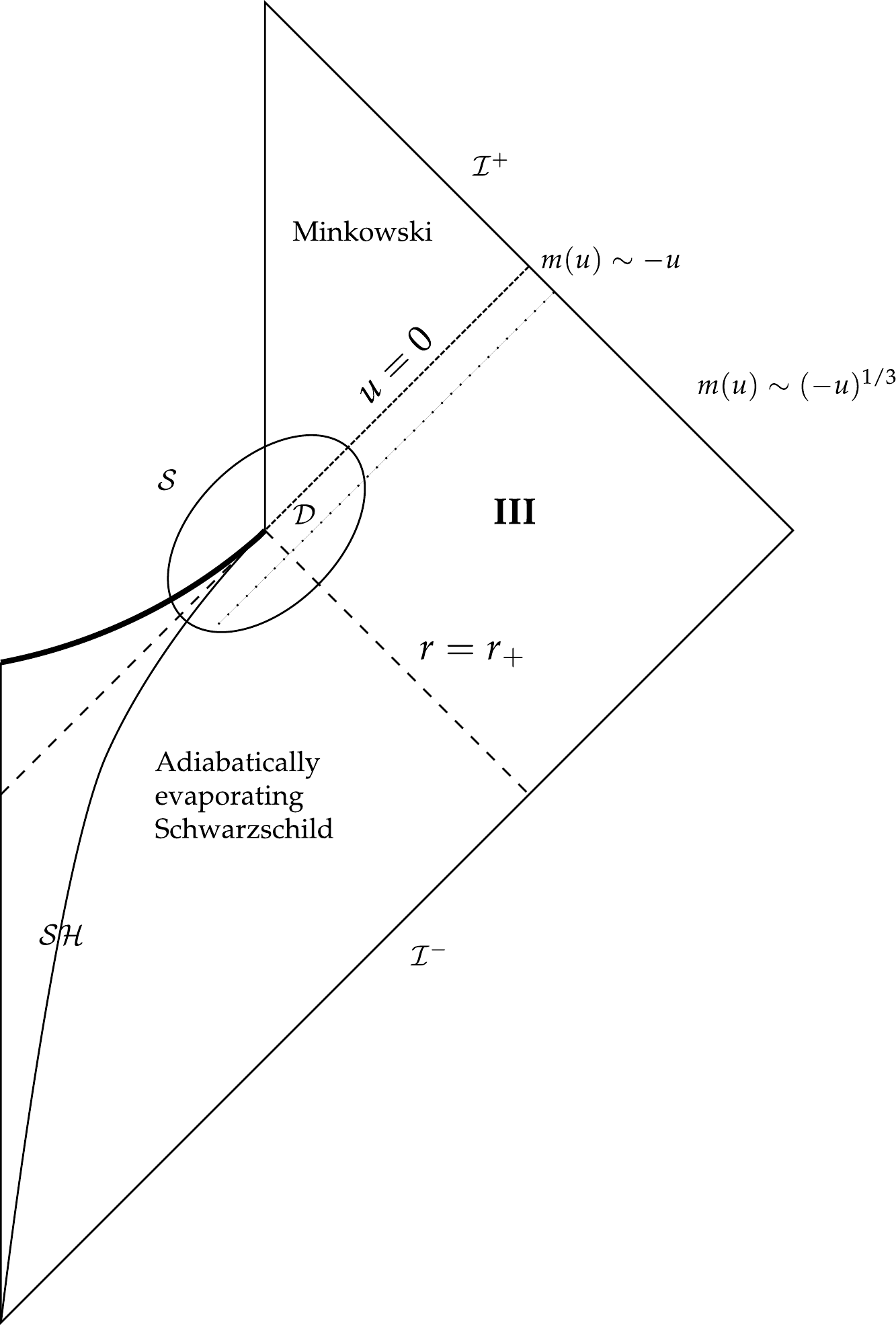}
\caption{An adiabatically evaporating Schwarzschild space-time 
evolving to a linear mass
Vaidya metric. $\mathcal{D}$ is the vanishing point and $\mathcal{SH}$ 
the stretched horizon. The region $\mathcal{S}$ is blown up in 
figure(\ref{fig:matching}).}
\label{fig:vaidyaevaporation}
\end{figure}

The motivation for this paper is to propose these linear-mass Vaidya metrics
as semi-classical models for the end-point of black hole evaporation. 
To make this more realistic, one should use a more general mass function
that follows the Hawking radiation formula $m(u) = (-u)^{1/3}$ up to the 
point that the mass of the black hole becomes Planckian, after which 
it becomes linear. It is generally 
hypothesised that black hole evaporation concludes with the complete
disappearance of the black hole and the singularity that it contains. 
We do not know precisely what happens beyond the future
horizon of this (possibly cataclysmic) event but one expects that the 
space-time returns to an asymptotically flat and non-singular configuration
(protected from more singular fates by cosmic censorship and positive 
mass theorems).
We will assume that the linear mass Vaidya metric is continuously attached 
to Minkowski space-time as in \cite{unruh} (some discussion of these 
issues in a slightly more general context can also be found in \cite{podolsky}). 
A qualitative sketch of this evolution is illustrated in 
figure (\ref{fig:vaidyaevaporation}).

To investigate the viability of this picture in this paper we perform
a detailed study of null geodesics and massless scalar fields
with non-zero angular momentum 
in the linear mass Vaidya metrics. We  will use a mixture of analytic and 
numerical techniques to get a complete picture of their behaviour. Our aim 
is to demonstrate that, close to the vanishing point, the null-singularity 
effectively becomes repulsive and consequently stable under the effects of
 small perturbations, backscattering and subsequent backreaction. This 
stability is a necessary condition for the suitability of this metric 
as a semi-classical model for the end-point of black hole evaporation. 

Most of this article is dedicated to the presentation and discussion of 
the analysis for massless particles with non-zero angular momentum.
In the first section we will look at null ingoing geodesics and then 
in the second section we study the wave-equation for a massless scalar in this 
background. In the concluding section we will present a discussion of
our results and some additional
evidence for our proposal that the linear mass Vaidya metric is a viable
candidate for a semi-classical model of the final stages of black hole
evaporation. 

\section{Null geodesics}

In the following we will be investigating particles in the background
of the linear mass Vaidya metric with $m(u) = -\mu u$ studied in detail 
by Waugh and Kayll Lake \cite{waugh861} 
\begin{equation}
\dd s^2 =
  -\left(
    1 + \frac{2\mu u}{r}
  \right)\dd u^2 -
  2\,\dd u\,\dd r +
  r^2\dd\Omega^2.
\end{equation}

For $0<\mu<1/16$ this class of space-times has the conformal structure 
illustrated in figure (\ref{fig:linearMvaidya}) -- note in particular that to 
the future of the vanishing point $\mathcal{D}$ there is no longer
a singularity. This metric (for $u<0$) is a solution to the Einstein equations 
with the stress-energy tensor
\begin{equation}
T_{\alpha\beta} = \frac{\mu}{4\pi Gr^2}\delta_\alpha^u\delta_\beta^u
\end{equation}
corresponding to a purely out-going spherically symmetric flux of 
radiation.

We are interested in null geodesics with non-zero angular momentum
in region II of figure \ref{fig:linearMvaidya}. For zero angular momentum the 
null geodesic equations are integrable and one can find various discussions 
of these in the literature e.g. \cite{waugh862}. 
The above coordinate system is however not very useful for the 
study of the metric in particular near the null singularity due to the 
degeneracy between $r$ and $u$ for $u\rightarrow 0^-$. 
In fact, as the singularity is approached we find that away from the 
end-point $r\rightarrow -u(1-\Delta)/4$ 
while precisely at the end-point $r\rightarrow -u(1+\Delta)/4$, where 
$\Delta = \sqrt{1-16\mu}$. It will be convenient to sometimes also use
the coordinate $z=-u/r$ in terms of which $r\rightarrow r_\pm = -u(1\pm\Delta)/4$, 
corresponding to $z\rightarrow z_\mp = (1 \pm \sqrt{1-16\mu})/4\mu$ 
as indicated in figure (\ref{fig:linearMvaidya}). Note that region II 
corresponds to $r_-<r<r_+$ ($z_+>z>z_-$, as $z_\pm = -1/r_\mp$).
To study the behaviour of the metric around the vanishing point ${\mathcal D}$
it is much more convenient to change to the double-null coordinates 
\cite{waugh861} for which the metric has the form
\begin{equation}
\dd s^2 = - 2 f(u,v)\, \dd u\, \dd v + r(u,v)^2 \dd\Omega^2
\label{doublenull}
\end{equation}
with source
\begin{equation}
T_{\alpha\beta} = \frac{\mu}{4\pi Gr(u,v)^2}\delta^u_\alpha\delta^u_\beta
\end{equation}
and where
\begin{equation}
f(u,v) = \frac{1 + \Delta}{2\Delta r(u,v)}
\left(r(u,v) + u(1-\Delta)/4\right)^{2/(1+\Delta)}.
\label{fuv}
\end{equation} 

In these coordinates $v=0$ coincides with $r=r_+=-u(1+\Delta)/4$ ($z=z_-$) while 
the singularity is at $u=0$ and $v<0$ corresponding to $r=r_-=-u(1-\Delta)/4$
($z=z_+$).
The coordinate transformation from $(u,r)$ to $(u,v)$ can be obtained, 
in principle, by solving the implicit equation \cite{waugh861}
\begin{equation}
|v|^{1 + \Delta}\left(r(u,v) + u(1-\Delta)/4\right)^{1-\Delta}
=\left(r(u,v) + u(1+\Delta)/4\right)^{1+\Delta}
\label{WKLradius}
\end{equation} 
for $r(u,v)$.
For the specific values $\Delta=1/5,1/3$ and $1/2$
where $\Delta = \sqrt{1-16\mu}$, $r(u,v)$ can be found explcitly by solving
at most a cubic polynomial equation\footnote{One could also go to quartic 
polynomials, for $\Delta = 1/7$ and $\Delta=3/5$ but the general behaviour 
of solutions does not change so we will not pursue these more complicated 
solutions
further.}, leading to the following expressions
for $r_\Delta(u,v)$;
\begin{eqnarray}
r_{1/5}(u,v)  =&  \frac {1} {30} \left (\sqrt[3] {5} v\left (\sqrt[3] 
{-\sqrt {27\left (u^3\left(27 u - 40 v^3 \right) \right)} +
27 u^2 - 180 u v^3 + 200 v^6} 
\right. \right. \nonumber \\ 
+& \left. \left. \sqrt[3] {\sqrt {27\left(u^3\left (27 u - 
40 v^3 \right) \right)} + 27 u^2 - 180 u v^3 + 
200 v^6}\right)  - 9 u + 10 v^3 \right),
\label{r15}
\end{eqnarray}
\begin{equation}
r_{1/3}(u,v) = \frac{1}{2}\left (v\sqrt {v^2 - \frac {2 u} {3}} 
- \frac {2 u} {3} +   v^2 \right),
\label{r13}
\end{equation}
\begin{equation}
r_{1/2}(u,v) = 
\frac {1} {8}\left (\frac {4} {3^{2/3}}\left(\sqrt[3] 
{\sqrt {3 v^6\left (27 u^2 - 64 v^3 \right)} - 9 u v^3} + \sqrt[3] 
{\sqrt {3 v^6\left (27 u^2 - 64 v^3 \right)} + 
9 u v^3} \right) - 3 u \right).
\label{r12}
\end{equation}
These explicit forms for $r(u,v)$ will be used for the quantitative
numerical analysis of the geodesic equations and below also 
for the numerical study of a massless scalar field on this geometry.

Due to the spherical symmetry angular momentum $L$ is conserved and
the null condition is
\begin{equation}
\frac{L^2}{r^2} = \left(1 + \frac{2\mu u}{r}\right)\dot{u}^2
+ 2\dot{u}\dot{r} = 2 f(u,v) \dot{u}\dot{v}.
\end{equation}
In addition, the linear mass Vaidya metrics have a homothety 
symmetry under rescalings of $u$ and $r$ with a corresponding overall 
rescaling of the metric. This gives rise to an additional conserved quantity
\begin{equation}
P = \left(1 + \frac{2\mu u}{r}\right)u\dot{u} + r\dot{u} + u\dot{r}
\end{equation}
Solving these two equations for $\dot{u}$ and $\dot{v}$ we find
\begin{equation}
\dot{u}^\pm = P\frac{1\pm  
\sqrt{1-\frac{L^2}{P^2}\frac{u}{r}k(u/r)}}
{rk(u/r)}
\end{equation}
and
\begin{equation}
\dot{v}^\pm = \frac{L^2}{2fr^2\dot{u}^\pm},
\end{equation}
where
\begin{equation}
k(u/r) = 2\mu\frac{u^2}{r^2} + \frac{u}{r} + 2.
\end{equation}

\begin{figure}[ht!]
\begin{center}
\subfigure[$\dot{u}^+(u,v)$]{
\label{fig:udotminus}
\includegraphics[width=0.45\textwidth]{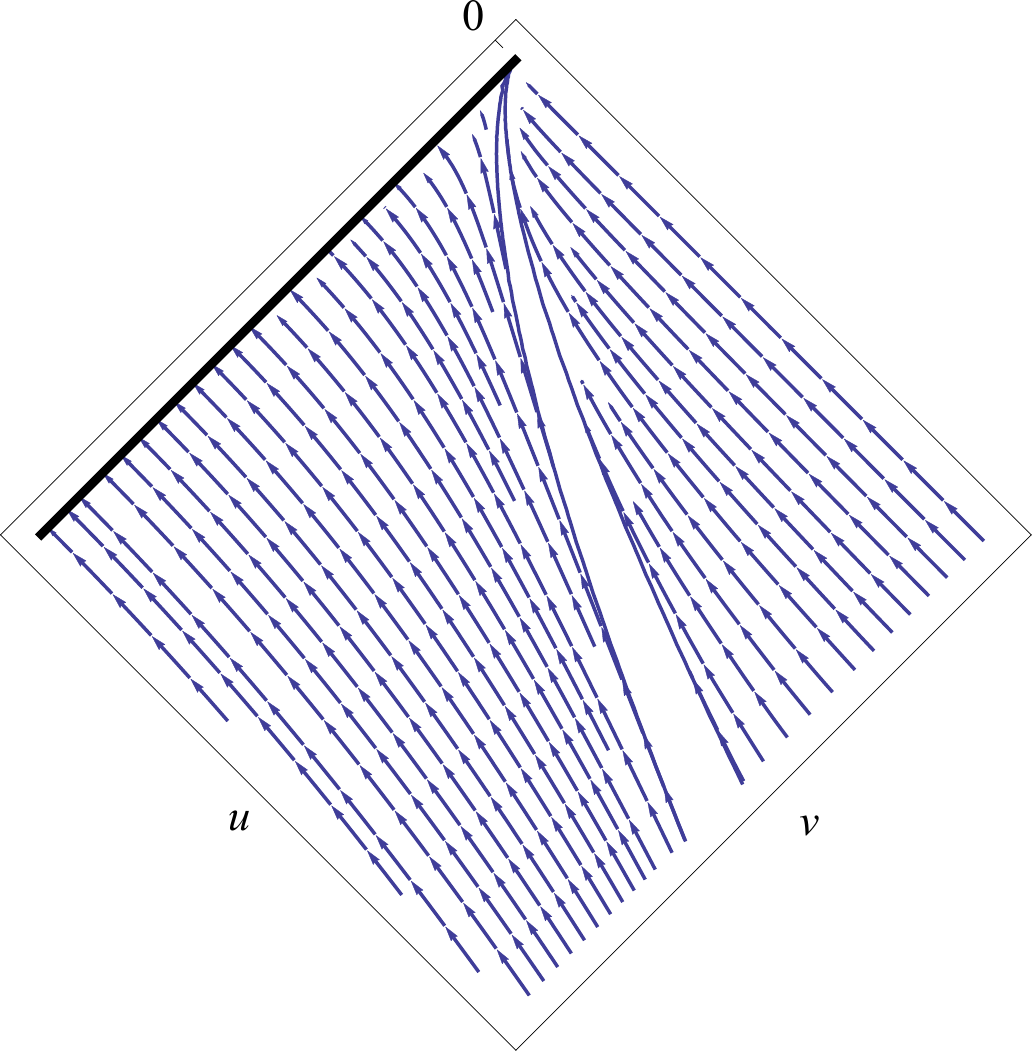}
}
\subfigure[$\dot{u}^-(u,v)$]{
\label{fig:udotplus}
\includegraphics[width=0.45\textwidth]{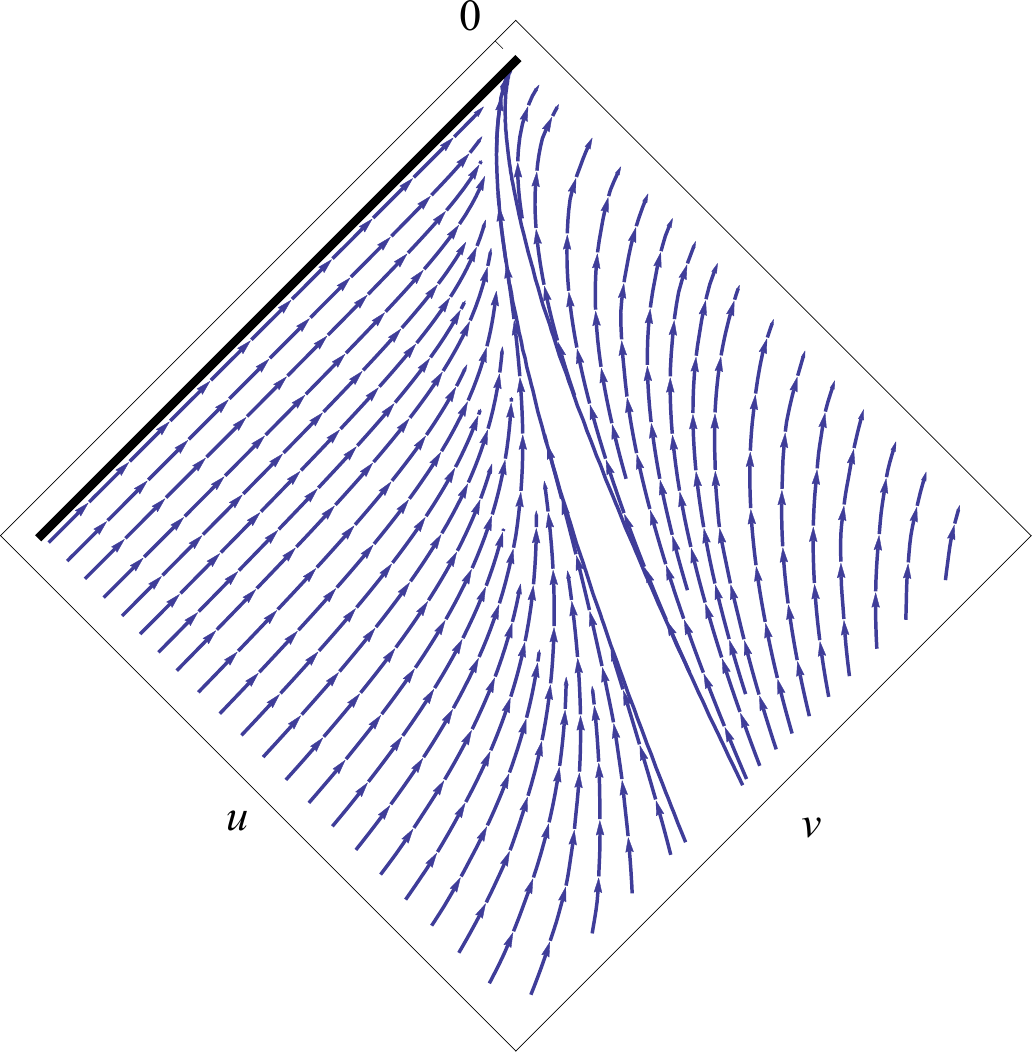}
}
\end{center}
\caption{Region II near the singularity for $\Delta=1/2$. The white region is
classically forbidden when $P^2/L^2 < h_{\text{max}}=3.75575$ and for the plot we
chose $P^2/L^2 = 3.7$.}
\label{fig:udot}
\end{figure}

In region II, between the two roots $r=r_\pm$ of $k(u/r)$
we demand that both $\dot{u}$ and $\dot{v}$ are positive (guaranteeing that 
the geodesics are future directed) and this 
leads to the requirement that $P<0$. This implies that to each 
value of $P<0$ there correspond two distinct solutions 
for $(\dot{u}^\pm,\dot{v}^\pm)$. 
They are both physical but correspond to two different ranges of 
initial conditions. We can qualitatively understand the behaviour of these 
solutions by considering the argument of the square root in the expression 
for $\dot{u}$. Clearly any complete classical trajectory (that 
does not run into the singularity) should be such that the 
argument of the square root is always positive, leading to the constraint that
\begin{equation}
h(z) = - z(2\mu z^2 - z + 2) \leq \frac{P^2}{L^2}.
\end{equation}
The crossover from the $+$ branch solutions for $(\dot{u},\dot{v})$ to the 
$-$ branch solutions occurs at the trajectory $h(z) = P^2/L^2$. 
These special trajectories are indicated in figure (\ref{fig:udot}) by 
the boundaries of the classically forbidden regions.

$h(z)$ has a maximum at $z=z_{\text{max}}$ between $z^+$ and $z^-$ 
along the curve
\begin{equation}
\frac{u}{r} = - z_{\text{max}}  = - \frac{1}{6\mu}(1 + \sqrt{1-12\mu})
\end{equation}
and its value there is
\begin{equation}
h_{\text{max}}=\frac{(1 + \sqrt{1-12\mu})(1 + \sqrt{1-12\mu}-24\mu)}{108\mu^2}.
\end{equation}

Ingoing particles with $z<z_{\text{max}}$ and $P^2/L^2<h_{\text{max}}$ will be 
reflected by the potential barrier continuing 
on to $\mathcal{I}^+$. Taking the limit of small $\mu$, 
one finds the bound $P^2/L^2<1/(27\mu^2)$ which is the same as that 
for massless geodesics outside a Schwarzschild black hole of mass $\mu$. 
For particles with $P^2/L^2>h_{\text{max}}$ on the other hand, they clearly 
run into the singularity coming in from $z<z_{\text{max}}$. The numerical
plots of the vector fields $(\dot{u}^\pm,\dot{v}^\pm)$ shown in
figure (\ref{fig:udot}) 
confirm this qualitative analysis. We will refer to the curve
$z=z_{\text{max}}$ as the Vaiyda photon sphere. 
It plays exactly the same role
as the photon sphere for the Schwarzschild metric which lies at
$r= 3GM$ and has a height $E^2/L^2 = 1/(27 M^2)$ -- incoming/outgoing photons 
below the barrier will be reflected while those above the barrier continue
in the same direction.

This above analysis gives us a clear picture of the qualitative behaviour
of geodesics, however 
the full numerical analysis of these equations remains quite complicated.
To confirm our qualitative discussion, we can extract some 
analytic information from the geodesic equations if we 
take a near singularity limit. In particular we 
can expand around small $v$, studying the behaviour as $u\rightarrow 0^-$.
The ratio $\dot{u}/\dot{v}$ expanded for small $u$ and $v$
\begin{equation}
\frac{\dd u}{\dd v} = \frac{\dot{u}}{\dot{v}} = (-u)^a(-v)^b
\end{equation}
is characterised by the exponents $a$ and $b$ with solution
\begin{equation}
u(v) = - \left(\frac{1-a}{1+b}(-v)^{1+b} + C\right)^{1/(1-a)}.
\end{equation}
We can easily see that geodesics will avoid the singularity at $u=0$ and
$v<0$ only if both $1-a$ and $1+b$ are negative.
To determine the exponents we need an expansion of $r(u,v)$ 
around $u=0$. In the region $v<0$ and for $u\rightarrow 0^-$ one can solve
to sub-leading order in $u$ the implicit equation \eqref{WKLradius} 
for $r_\Delta(u,v)$ giving
\begin{equation}
r_\Delta(u,v) = -\frac{u(1-\Delta)}{4} 
+ \left (-\frac{\Delta u}{2 v}\right )^{(1+\Delta)/(1-\Delta)}.
\end{equation}
It is easy to check that for the $(u^+,v^+)$ branch
$a=(1-3\Delta)/(1-\Delta)$ and $b=2\Delta/(1-\Delta)$, while for the 
$(u^-,v^-)$ branch we have $a=(1+\Delta)/(1-\Delta)$ and $b=-2/(1-\Delta)$.
Thus for the $+$ branch there is a class of geodesics that can 
run into the singularity while for the $-$ branch the geodesics go 
precisely to the vanishing point at $u=v=0$. 
Alternatively we can look at an expansion relevant for geodesics
approaching the potential barrier from the external region. The ``$-$'' branch
is such that all geodesics cross the $v=0$ line before reaching $u=0$
while all of the $+$ branch solutions again go to the vanishing point at
$u=v=0$. These conclusions clearly agree with the flow lines of 
the explicit solutions for $(\dot{u}^\pm,\dot{v}^\pm)$ displayed
in figure (\ref{fig:udot}).

\section{Massless scalar field}

To further understand the nature of the null singularity and the potential 
barrier that surrounds it we will now present the results of the numerical
analysis of a massless scalar 
field in this geometry, in particular studying the scattering of 
an in-going gaussian wave packet from the near singularity 
region. 

\begin{figure}[ht!]
\begin{center}
\subfigure[$\Delta = 1/2$]{
\label{fig:VVaidya12}
\includegraphics[width=0.45\textwidth]{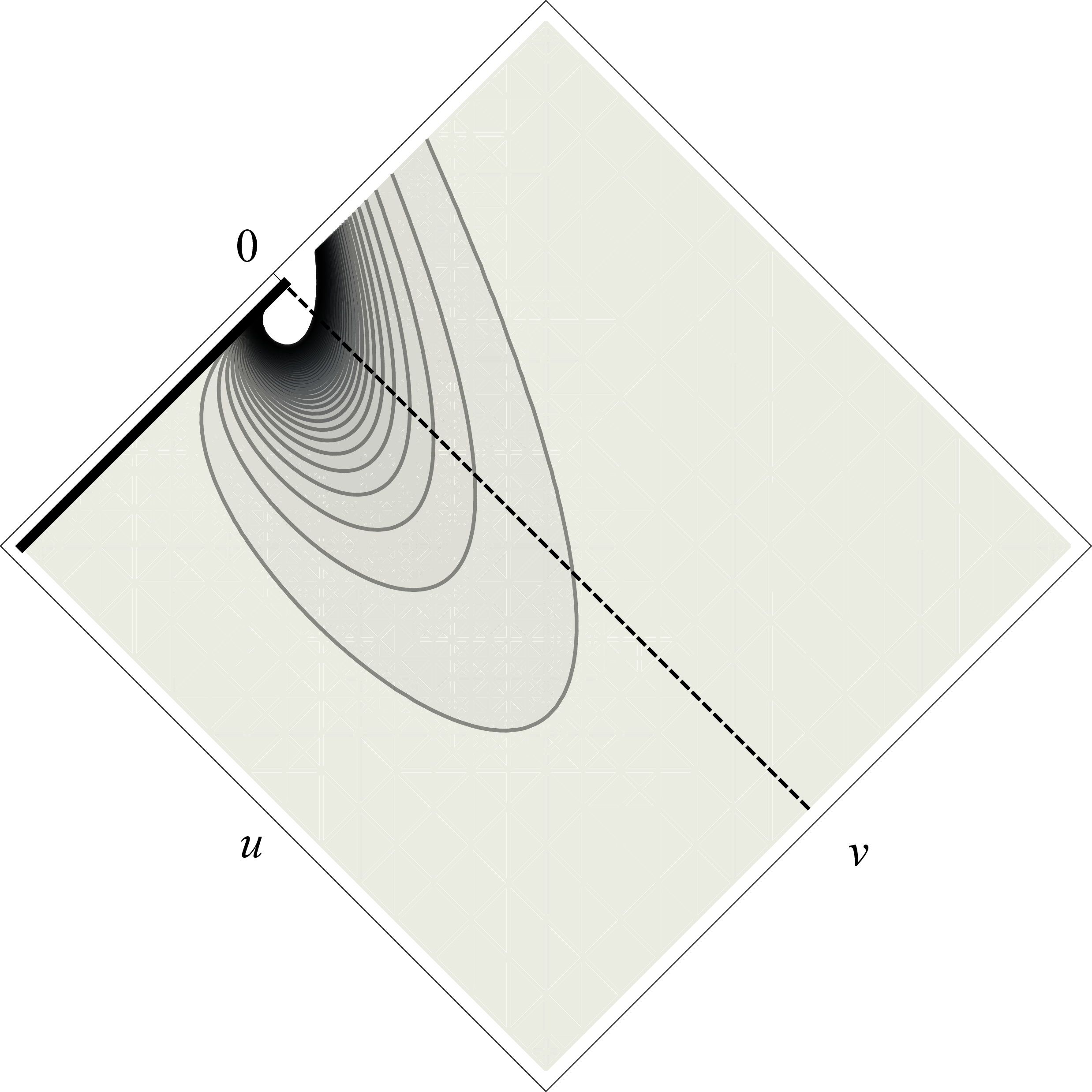}
}
\subfigure[$\Delta = 1/5$]{
\label{fig:VVaidya15}
\includegraphics[width=0.45\textwidth]{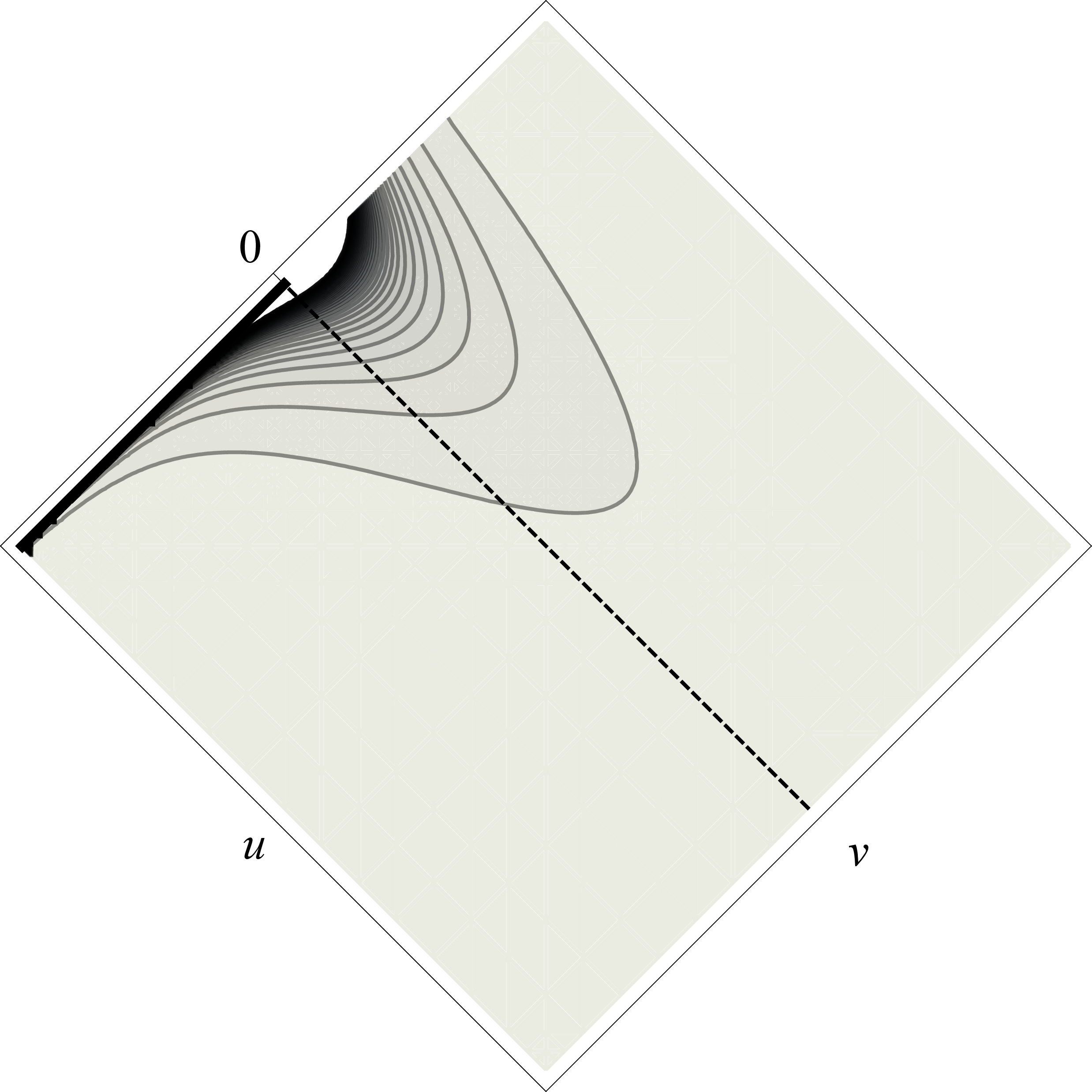}
}
\end{center}
\caption{Scattering potentials on the boundary between regions II and 
III. Darker shades correspond to larger $V(u,v)$ with divergence 
precisely at $u=v=0$ for $\Delta=1/2$ and at $u=0$, $v<0$ for 
$\Delta = 1/5$.}
\label{fig:VVaidya}
\end{figure}

In double-null coordinates the wave-equation for a massless scalar
field is
\begin{equation}
\frac{\partial^2\Psi}{\partial u\,\partial v} 
+ \frac{\ell(\ell+1)}{2r^2}f\Psi = 0.
\label{waveqn}
\end{equation}
The potential $V(u,v)$ is infinite around the vanishing 
point, and is also divergent along the null-singularity when $\Delta<1/3$
as can be seen clearly in figure (\ref{fig:VVaidya}).  As can be
seen from the plots the potential for $\Delta=1/2$ grows more rapidly 
in the region $v<0$ than does that for $\Delta=1/5$, the crossover
between these two behaviours is at $\Delta=1/3$ (not shown) for 
which the potential is completely symmetric about $v=0$. 

Thus to understand the physics of scalar and other particles near the 
singularity we simply need to study $V = \ell(\ell+1)f/2r^2$.
Taking the expression for $f(u,v)$ from \eqref{fuv} we obtain,
in the limit that $u\rightarrow 0$ while $v<0$, the leading term in $V(u,v)$,
\begin{eqnarray}
V(u,v) & =& \ell(\ell+1)\left (\frac{1+\Delta}{2\Delta}\right )
\left(-\frac{4}{u(1-\Delta)}\right)^3
\left(-\frac{\Delta U}{2v}\right)^{2/(1-\Delta)}\nonumber \\
& = &\lambda(\Delta)(-u)^{(-1+3\Delta)/(1-\Delta)}(-v)^{-2/(1-\Delta)}
\end{eqnarray}
where
\begin{equation}
\lambda(\Delta) \sim \Delta^{(1+\Delta)/(1-\Delta)}\frac{(1+\Delta)}{(1-\Delta)}.
\end{equation}

As the potential near the singularity is a simple product of a function 
of $u$ and a function of $v$ we can make the ansatz that 
near the singularity the wave-function factorizes 
$\Psi(u,v) \sim U(u)V(v)$ and thus \eqref{waveqn} 
can be trivially integrated to obtain
\begin{equation}
\Psi(u,v) = \mathcal{N}e^{\kappa(u,v)},
\end{equation}
where the exponent $\kappa(u,v)$ is
\begin{equation}
\kappa(u,v) = \frac{2C\Delta}{1-\Delta}(-u)^{2\Delta/(1-\Delta)}
+\frac{\lambda(\Delta)}{C}
\frac{(1-\Delta)}{(1+\Delta)}(-v)^{-(1-\Delta)/(1+\Delta)}.
\end{equation}
To avoid possibly unphysical divergences at $v\rightarrow 0^-$ we require that
the integration constant $C>0$. This means that the singularity is mildly
repulsive and thus is not sufficient to completely repel an incoming 
particle. However it does become more repulsive as $v\rightarrow 0^-$. 

Looking instead at the behaviour 
of the potential for small $v$, along the line $v=0$ ($z=z_-$) we find that the 
leading term in \ref{WKLradius} for small $v$ is simply 
\begin{equation}
r(u,v) = - \frac{u(1+\Delta)}{4},
\end{equation}
giving the leading term in the potential
\begin{equation}
V(u,v) \approx \kappa u^{-(1+3\Delta)/(1 + \Delta)}.
\end{equation}
Inserting this into the wave-equation we find as the leading 
behaviour of the wave-function along the line $z=z_-$
\begin{equation}
\Psi(u,v) \sim \exp\left(-u^{-2\Delta/(1+\Delta)}\right),
\label{psiv0}
\end{equation}
which clearly goes to zero for $0<\Delta<1$. Ingoing waves 
localised around $z_-$ are thus repelled from the singularity
as anticipated above. The effectiveness of 
this repulsion is obviously greater as $\Delta\rightarrow1$.

\begin{figure}
\centering
\includegraphics[width=0.45\textwidth]{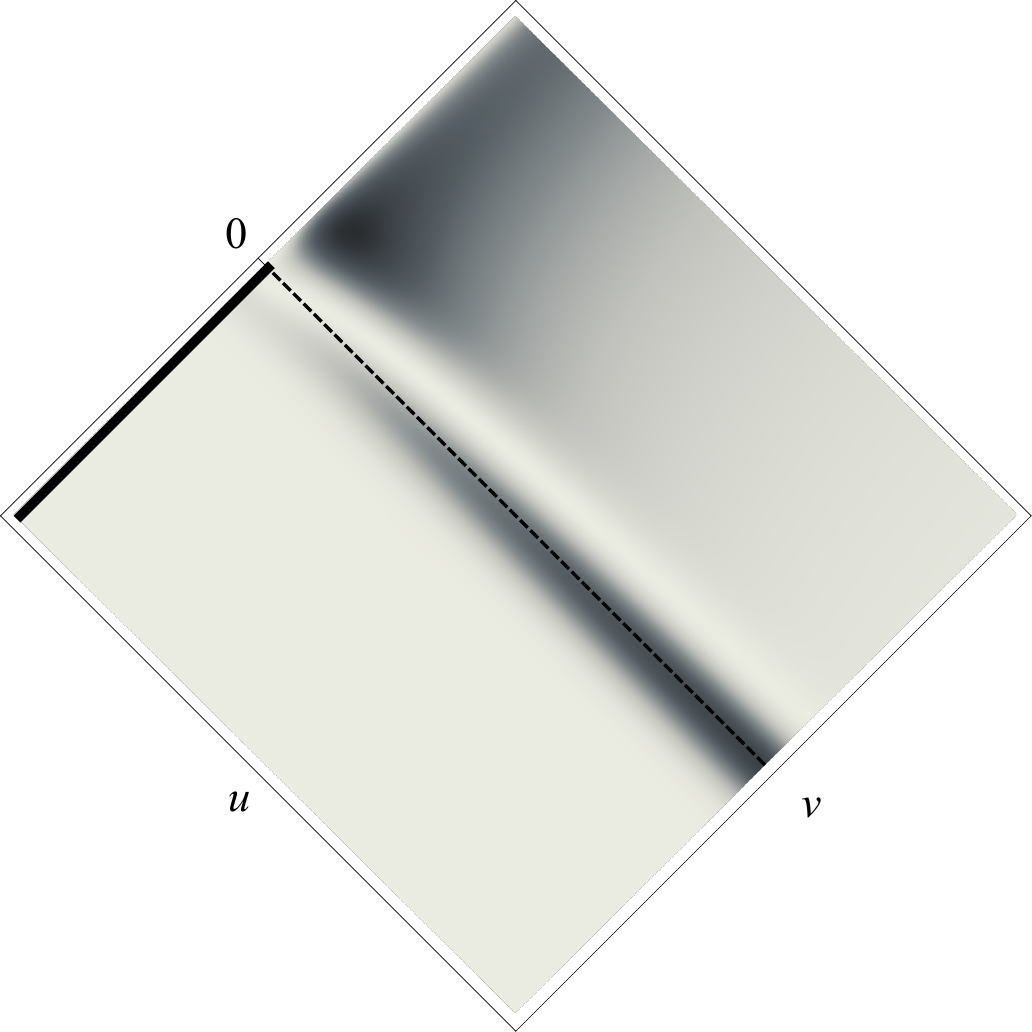}
\caption{Plot of $|\Psi(u,v)|^2$ showing the scattering of a Gaussian 
wave-packet with $\ell=1$ and centre at $v=0$ from the 
singularity of the $\Delta=1/2$ metric. Light-shaded regions correspond to 
$|\Psi|^2 \approxeq 0$.}
\label{fig:scattering}
\end{figure}

To check the global consistency of this analysis and the detailed 
behaviour close to the singularity we once again used the 
explicit solutions for $r_{1/2}(u,v)$ of \eqref{r12} 
to carry out the numerical integration of 
the wave-equation on the linear mass Vaidya metric for $\Delta = 1/2$. 
The integration method used was the double-null characteristics 
technique as originally proposed in \cite{gundlach}. The incoming wave-function
is a gaussian and boundary conditions at $v=v_0<0$ were simply 
$\Psi(u,v_0) = 0$, again as proposed and used in \cite{gundlach}. The results
for $\Delta=1/2$ are shown in figure (\ref{fig:scattering}), the results are 
very similar to those for $\Delta = 1/5,1/3$. It is clear that the 
behaviour is exactly that predicted in the above calculations. Note 
in particular that ingoing waves very close to the singularity and 
with $v\lesssim 0$ are strongly reflected towards $\mathcal{I}^+$. 

\section{Discussion}

This final section will be a speculative discussion about the possible
role of the linear Vaidya metric in the final stages of black hole
evaporation. It will necessarily be much more qualitative than the preceeding
sections, but we believe that the general semi-classical 
picture for the final stages of black hole evaporation should have 
features similar to that which we will present here. 

\begin{figure}
\centering
\includegraphics[width=0.7\columnwidth,keepaspectratio=]{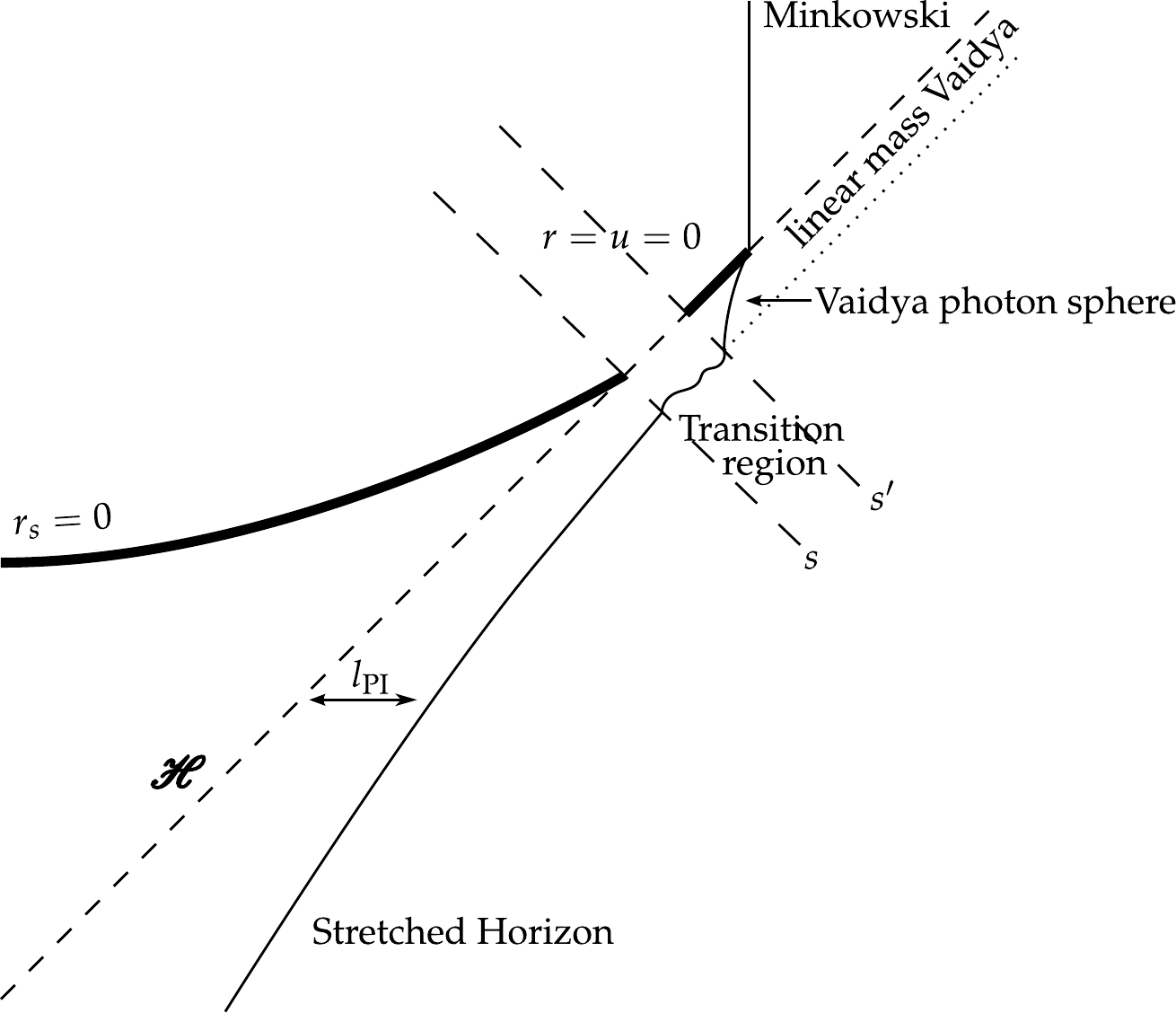}
\caption{A blowup of the region $\mathcal{S}$ of figure 
(\ref{fig:vaidyaevaporation}).}
\label{fig:matching}
\end{figure}

As we do not have control over the late stages of the evolution of 
the Schwarzschild black-hole we will assume that there is a Planckian sized
transition region between the lines $s$ and $s^\prime$ in figure 
(\ref{fig:matching}), 
interpolating from a near Planck-mass Schwarzschild black hole
to a linear mass Vaidya metric with $0<\mu<1/16$.
To investigate the possible consistency of this proposal 
we will first look at the Kretschmann 
scalar as a measure of the magnitude of the curvature.

For a Schwarzschild black hole of mass $M$  
\begin{equation}
K_{\text{S}} = \frac{48 G^2 M^2}{r^6}
\end{equation}
while for Vaidya
\begin{equation}
K_{\text{V}} = \frac{48 \mu^2 u^2}{r^6}.
\end{equation}
We propose that the linear mass Vaidya metric should be inserted at a point 
in the evaporation where the stretched 
horizon at $r=G(2M+M_{\text{Pl}})$ meets the photon sphere of the Vaidya metric
at $z=z_{\text{max}}$  and at this point the strength of the 
curvatures of Schwarzschild and Vaidya should be similar. Furthermore, continuity
of the radius of the transverse sphere requires that the radial coordinates
in Schwarzschild and Vaidya coincide in the matching region $\mathcal{T}$. 
Looking at the Kretschmann scalar we thus require that
\begin{equation}
\frac{M}{2M+M_{\text{Pl}}}=\mu z_{\text{max}}=\frac{1}{6}(1 + \sqrt{1-12\mu}).
\end{equation}
Introducing mass measured in Planck units $\bar{M}=M/M_{\text{Pl}}$ we find
\begin{equation}
\frac{6\bar{M}}{2\bar{M}+1}=1 + \sqrt{1-12\mu},
\end{equation}
and this implies that
\begin{equation}
\mu=\frac{\bar{M}(1-\bar{M})}{(2\bar{M}+1)^2}.
\label{KMatching}
\end{equation}

In addition we would also like the heights of the photon sphere barriers to 
be similar in the cross-over region and thus
\begin{equation}
\frac{1}{27\bar{M}^2} = h_{\text{max}}(\mu),
\end{equation}
or solving for $\mu$,
\begin{equation}
\mu = -9\bar{M}^2 - 216\bar{M}^4 + \sqrt{\bar{M}^2 + 108 \bar{M}^4 + 
3888\bar{M}^6 + 46656\bar{M}^8}.
\label{PSMatching}
\end{equation}
Solving \eqref{KMatching} and \eqref{PSMatching} simultaneously 
gives $\bar{M} = 0.5086 \approx 1/2$ at $\mu = 0.0614 <1/16$,
indicating that the ideal matching point is close to $M=M_{\text{Pl}}/2$. 
If we consider only the matching of the Kretschmann scalar we find 
that we are restricted to a crossover from Schwarzschild to Vaidya 
that takes place in the interval $M_{\text{Pl}}/2 < M < M_{\text{Pl}}$
corresponding to $0<\mu<1/16$.  However, from (\ref{KMatching}), 
we see that increasing the mass at the transition region towards 
$M=M_{\text{Pl}}$ corresponds to $\mu\rightarrow 0$. In this limit we find that
the height of the photon barrier of Schwarzschild decreases as 
$1/(27 \bar{M}^2)\rightarrow 1/27$ whereas that of the Vaidya metric 
increases as $1/(27\mu^2)\rightarrow\infty$.
Note also that, from the discussion of the previous section, the potential 
for scattering near the singularity and with $v<0$ is more repulsive 
when $\Delta\rightarrow 1$ ($\mu\rightarrow 0$)
as can be seen from \eqref{psiv0} and in figure (\ref{fig:VVaidya}).
If indeed the matching occurs closer to $M_{\text{Pl}}$ it is necessary that 
after the Page time \cite{page}, when the mass of the black 
hole is on the order of several times $M_{\text{Pl}}$, 
the space-time becomes modified such that the potential rises more rapidly.
The ideal choice of transition region requires some more information 
about the evolution after the Page time. 

At this point we would like to add a couple of comments on the stability 
and self-consistency of our construction. In \cite{waugh862}
it has been argued that the null singularity of these metrics may be unstable to 
backscattering due to large blueshifts as particles approach 
the singularity. Back-scattered particles outside the photon sphere
will generically have $\ell\neq 0$
and thus will feel the repulsive nature of the singularity possibly 
reducing the effect of potential blue-shifts. 
Furthermore, outgoing particles that come from the near horizon region can 
escape to infinity only if they can pass the photon sphere from the inside,
implying that those that get out must have either an energy 
that is significantly transplanckian, or  zero angular momentum. In turn
this appears to imply that in the final stages of evaporation
outgoing Hawking radiation will predominantly have 
angular momentum close to zero leading to an a posteriori justification for 
the use of the Vaidya metric and for its stability. 

Another interesting effect of the linear mass Vaidya modification of the final 
phase of evaporation is a slight cooling down of the black hole to
$T\sim 1/\sqrt{M}$ compared to the well-known $T\sim 1/M$ of Schwarzschild. 
This is still divergent for vanishing mass, but it should be noted that 
in \cite{balbinot} the inclusion of backreaction causes a more significant
cooling down with a finite temperature at the end-point of evaporation.

Obviously this is not a complete analysis of the cross-over region but 
we believe that the picture we are presenting is quite plausible. 
To make this more concrete it would be interesting to search for 
a deformation of the linear mass Vadiya metric for which the 
stress energy tensor is only spherically symmetric in the final 
Planckian region and becomes more like the stress-energy tensor 
of Hawking radiation further in the null past, thus providing also 
a semi-classical model for the adiabatically evaporating Schwarzschild 
to linear mass Vaidya transition. Along these lines there is a proposal 
in \cite{podolsky} for an evolution from a Robinson-Trautmann metric 
to a Vaidya metric in the final phase of evaporation. 

Finally it is worthwhile highlighting the appearance 
of a scale-invariant metric in the final stages of evaporation. It would 
be very interesting to further investigate the possible role of scale invariance 
in black hole production and evaporation. Some interesting examples in black hole
physics where scale invariance can be found are Choptuik scaling 
\cite{choptuik} and the universality of scale-invariant metrics in the 
Penrose limit of space-like and null space-time singularities \cite{blau}.

{\bf Acknowledgements}
I would like to thank M. Blau and D. Veberi\v{c} for useful comments 
and discussions. I would also like to thank A. Su\v{s}nik for help with 
the figures.

\end{document}